\documentclass{article}

\def\beq{\begin{equation}}
\def\eeq{\end{equation}}
\def\pdr{\partial}

\def\m#1{$#1$}
\def\half{{1\over 2}}
\usepackage{amssymb}
\def\beqs{\begin{eqnarray}}
\def\eeqs{\end{eqnarray}}
\def\[{\left[}
\def\]{\right]}
\def\({\left(}
\def\){\right)}
\def\Lie{{\cal L}}

\begin{document}

\centerline{\bf A Hamilton-Jacobi Formalism for Thermodynamics}

\begin{center}{\sf S. G. Rajeev}\\
Department of Physics and Astronomy\\ 
Department of Mathematics\\
University of Rochester\\
  Rochester NY 14627
\end{center}


\centerline{\bf Abstract}
    
    We show that  classical thermodynamics has a formulation in terms of Hamilton-Jacobi theory, analogous to mechanics. Even though the thermodynamic variables come in conjugate pairs such as pressure/volume or  temperature/entropy, the phase space is odd-dimensional. For a system with \m{n} thermodynamic degrees of freedom it is \m{2n+1}-dimensional. The equations of state of a substance  pick out an \m{n}-dimensional submanifold. A family of substances whose equations of state depend on \m{n} parameters define a hypersurface of co-dimension one. This  can be described by the vanishing of a function which plays the role of a Hamiltonian. The   ordinary differential equations (characteristic equations)  defined by this function describe a dynamical system on the hypersurface. Its orbits can be used to reconstruct the equations of state. The `time' variable associated to this dynamics is related to, but is not identical to, entropy. After developing this formalism on well-grounded systems such as the van der Waals gases and the Curie-Weiss magnets, we derive a Hamilton-Jacobi equation for  black hole thermodynamics  in General Relativity. The cosmological constant appears as a constant of integration in this picture.

\pagebreak
\section{ Introduction}

General Relativity is surprisingly similar in many ways to Thermodynamics.  It predicts an irreversibility in  the dynamics of black holes: the area of the horizon must increase. Since we cannot see the interior of a black hole, they carry  an entropy which must be proportional to this  area.  The analogy suggests that GR is the average over some as yet unknown microscopic degrees of freedom.

The situation is analogous to that with the theory of gases at the end of the nineteenth century. It was clear that the kinetic theory of gases gave a good derivation of the ideal gas law and through the work of Boltzmann, a possible derivation of thermodynamic laws from the first principles of mechanics. Yet, it was not known then what the value of Boltzmann's constant was, or equivalently, how big molecules are. 
Yet , van der Waals was able to deduce a simple equation for the departure of a gas from ideal by simple considerations of molecular dyamics. In a similar way, it  might be possible to test microscopic theories of gravity by their implications on the thermodynamics of  black holes or of  the universe. We find already an intriguing connection between black hole thermodynamics and the cosmological constant.

 This motivation  led us to questions about thermodynamics that turned out to be of independent   interest . In particular, we found  a dynamical system on the thermodynamic phase space that is  analogous to classical mechanics but is still quite different. The `time' variable of this evolution of the thermodynamic variables is related to, but is not always the same as, entropy.  Although many of the ingredients of our theory were known in the nineteenth century, this particular form of thermodynamics appears to be still new. We have tried to make the paper self-contained by reviewing some of the background material on mechanics (section 1), thermodynamics of gases (section 2) and magnets  (section 3) and contact geometry (section 4). The erudite reader could skip ahead to sections 2.3, 3.3, 5 and 6 for the new material.

\subsection{ The Three Classical Disciplines}

Three important disciplines of classical physics are  Mechanics,  Ray Optics and  Thermodynamics. The work of Hamilton and Jacobi  showed that  Mechanics and Ray optics had a very similar structure: they can both  be formulated in terms of the Hamilton-Jacobi (or eikonal) equation. We now understand why that is so: they are both approximations to wave theories in the limit of short wave length.

Let us invert history and start with the  wave equations of optics and quantum mechanics\cite{BornWolf, LandauLifshitz}
\beq
{n^2(x)\over c^2}{\pdr^2\over \pdr t^2}\phi=\nabla^2\phi, \quad i\hbar {\pdr \psi\over \pdr t}=-{\hbar^2\over 2m}\nabla^2\psi+V(x)\psi
\eeq
If we put \m{\phi=e^{i S\over \lambda},  \psi=e^{iS\over \hbar}} these become in the limit \m{\lambda\to 0,\hbar\to 0}:
\beq
{n^2(x)\over c^2}\left({\pdr S\over \pdr t }\right)^2=(\nabla S)^2,\quad  
{\pdr S\over \pdr t} ={(\nabla S)^2\over 2m}+V(x) \eeq
These are  the Eikonal and  Hamilton-Jacobi equations. The phase of the wave  is constant on the wave front ( the surface \m{S=} constant)  and varies most rapidly along the vector \m{\nabla S} which is normal to the wavefront. The wave can be thought of as propagating along the ray that solves the ODE
\m{
{dx\over dt}=\nabla S.
}

\subsection{Hamilton's  Equations}

Thus \m{\nabla S} is proportional to the wavenumber (momentum).  We can identify also \m{\pdr S\over \pdr t} with frequency (energy) up to a constant. A Hamilton-Jacobi equation of the form
\beq
H\left(x,\nabla S
\right)={\pdr S\over \pdr t}
\eeq
is equivalent to the system of ordinary differential equations that determine the rays\cite{CourantHilbert2}
\beq
{dx\over dt}={\pdr H\over \pdr p} ,\quad {dp\over dt}=-{\pdr H\over \pdr x}.
\eeq
These in turn are equivalent to Newton's laws of mechanics ( or Snell's law of refraction).

Position and momentum appear symmetrically in the above equations. In fact there is a much bigger symmetry. Any observable  \m{F(x,p)}  generates a one-parameter  family of co-ordinate transformations through Hamilton's equations. These are the canonical transformations generated by that observable. Time evolution is simply the canonical transformation generated by the energy ( Hamiltonian) function. Momentum generates translations, angular momentum generates rotations etc. Observables come in conjugate pairs generating translations on each other\cite{LandauLifshitz}.

\subsection{Classical  Thermodynamics} 

In this paper we will show that  thermodynamics   is a third facet  of  this classical framework. Gibbs \cite{Gibbs} realized  that the basic law of thermodynamics, combining the first and second laws, has  a geometric meaning\cite{Gibbs}.  The theory of differential forms was developed by Pfaff, Caratheodory and others to give his  insight an elegant mathematical formulation.  Chandrashekhar \cite{Chandrashekhar} gives a rare treatment of this subject in a textbook. An analogy of thermodynamics with mechanics  and optics is    mentioned   occasionally \cite{thermobooks, Buchdahl, Peterson, Omohundro}.  The most detailed geometrical formulation is in the review article Ref. \cite{Arnold}.   It is shown there that a hypersurface in a contact manifold defines a dynamical system.

But the physical meaning of this dynamics appears not to be clarified anywhere\footnote{This analogy with mechanics was noticed in Ref. \cite{Peterson}, but it is not as perfect as implied there. In particular, because the thermodynamic phase space is odd dimensional (see below), the correct formalism uses contact geometry and Legendre transformation rather than symplectic geometry and Poisson brackets.}. What is the meaning of such a hypersurface in thermodynamics? It is certainly not the submanifold  defined by the equation of state of a material, as it has the wrong dimension. (More precisely,  it is not a Lagrangian submanifold. See  below).  What is the Hamilton-Jacobi equation of thermodynamics for some familiar systems like gases or magnets? What is the meaning of the `time' variable in this dynamics? What are the implications for the thermodynamics of black holes? These are the questions we will begin to address in this paper. After understanding these down to earth system we will turn to the dynamics of the simplest black hole.

\section{ Thermodynamics of Gases}
To be concrete let us consider  the most familiar thermodynamic system, a gas with a fixed number of molecules. It has five thermodynamic variables
\m{U,T,S,P,V}: the internal energy, temperature,
entropy, pressure and volume. \m{S,U,V} are {\em extensive variables}; i.e., are proportional to the size of the system. \m{P,T} are intensive: they do not scale with the size of the system. Two subsystems are in equilibrium if they have equal values for the intensive variables.

Extensive and intensive variables come in conjugate pairs such as pressure/volume or temperature/entropy. 
Yet there are always an odd number of variables.
This can be seen in the first law of thermodynamics which is a 
condition on the infinitesimal variations of these quantities:
\beq
\alpha\equiv dU-TdS+PdV=0.
\eeq
The conjugate pairs appear together as \m{PdV} or \m{TdS},except for internal energy which has no conjugate. 

An equivalent form would be
\beq
dS-T^{-1} dU-\tilde PdV=0
\eeq
where \m{\tilde P={P\over T}}.
Indeed any  variable  can be chosen to be in the privileged position as  the `fundamental variable' in the first term, with unity as the conjugate. This symmetry resolves the  apparent contradiction of having an odd variables, while at the same that variables
come in pairs. \m{\alpha} is only determined modulo multiplication by a non-zero function: only its kernel (zero set) has a physical significance.

Equivalent formulations of the first law are
\beq
d[U+PV]-VdP-TdS=0,\quad  d[U-TS]+PdV+SdT=0
\eeq
etc., in which the positions of some of the extensive and intensive variables are switched. These transformations are analogous to the canonical transformations of classical mechanics and are called Legendre transformations. More on this later.

\subsection{The Ideal Gas}

The particular two dimensional surface corresponding to each material is given by three equations of state to be satisfied among the variables \m{(T,P,s,u,v)}. For example, a monatomic 
ideal gas satisfies the laws
\beq
Pv=T,\quad u={3\over 2}T
\eeq
where \m{V=nRv} is the volume of  the gas, \m{U=nRu} the internal energy, \m{n} is the number of moles of the gas in the sample  and  \m{R\approx 8.3}JK$^{-1}$mol$^{-1}$ is the gas constant. The remaining relation needed to fix  the two dimensional surface of the ideal gas is obtained by integrating
the differential equation $\alpha=0$. This gives the entropy \footnote{\m{S=nRs}, where \m{n} is the number of moles of the gas and \m{R} the gas constant. We will find it useful similarly to define `specfifor all extensive variables \m{S=nRs,V=nRv,U=nRu,M=nRm} etc. The gas constant is put in to simplify later formulas.}:
 \beq
 ds -{P\over T}dv-{1\over T} du=0\Rightarrow  s=\log\left[{u^{3\over 2}v}\right].
 \eeq
 This holds up to an additive constant of integration; i.e.,   independent of \m{u,v}.  
 
 Conversely, once this `fundamental relation' between
 extensive variables is given the intensive variables
 are determined by 
 \beq
 {1\over T}=\left({\pdr s\over \pdr u}\right)_v,\quad {P\over T}=\left({\pdr
 s\over \pdr v}\right)_u
 \eeq
 which are the other equations of state.
 
\subsection{van der Waals Gases}

A simple model of a non-ideal gas is due to  van der Waals\cite{Sengers}:
\beq
P={T\over v-b}-{a\over v^2} \label{vdw1}
\eeq
where   \m{a,b} are constant parameters. We can regard \m{b} as the volume  excluded   because of the finite size of  the molecules. The constant \m{a} measures the strength of the short range attraction among the  molecules.

Let us now find the entropy:
\beqs
 ds&=&{P\over T}dv+{du\over T}\cr
&=& {dv\over v-b}+{du-{a\over v^2}dv\over T}
\eeqs
Thus \m{T} is an integration factor which must  make the last term an exact differential. A moments thought gives
\beq
T={2\over 3}\[u+{a\over v}\].\label{vdw2}
\eeq
We fix the constant  of integration by noting that the low density  limit \m{v\to \infty} must be the ideal gas. Thus,
\beq
s=\log\left[(v-b)\left(u+{a\over v}\right)^{3\over 2}\label{vdws}
\right]
\eeq
{\footnotesize To check:
\beq
{P\over T}=\left({\pdr s\over \pdr v}\right)_u={1\over v-b}-{3a
\over 2v^2}{1\over u+{a\over v}},\quad {1\over T}=\left({\pdr s\over \pdr u}\right)_v={3\over 2}{1\over u+{a\over v}}
\eeq
from which it follows that 
\beq
P={T\over v-b}-{a\over v^2}.\label{vdw3}
\eeq
}

Now we can eliminate the two parameters \m{a,b} from the three equations (\ref{vdw1},\ref{vdw2},\ref{vdw3}) to get an equation relating the thermodynamic variables:
\beq
  \[{P\over T} v-{1\over T}u+{3\over 2}\]^2T^{-3}={27\over 8}v^2e^{-2s}    \label{vdwhj1}
\eeq

This is the hypersurface in the thermodynamic phase space describing the whole family of van der Waals gases. We want to find such an equation for all the examples of thermodynamic systems we study.

\subsection{Hamilton-Jacobi equation for van der Waals gases}

From the equation (\ref{vdwhj1}) we can now recover the equations of state of the van der Waals gases. In the picture in which \m{s} is the thermodynamic potential and \m{u,v} are the co-ordinates we can put in 
for their conjugates 
\beq
T^{-1}=\({\pdr s\over \pdr u}\)_v,\quad {P\over T}=\({\pdr s\over \pdr v}\)_u
\eeq
to get the first order PDE
\beq
  \[ v{\pdr s\over \pdr v}-u{\pdr s\over \pdr u}+{3\over 2}\]^2\({\pdr s\over \pdr u}\)^{3}={27\over 8}v^2e^{-2s}.    
\eeq

This is the Hamilton-Jacobi equation of the thermodynamics of van der Waals gases. The complete integral  of this equation ( in the sense of Ref.\cite{CourantHilbert2}) will have two undetermined parameters. This is exactly (\ref{vdws}).

 \section{Magnetic systems}
 For a magnetic system, the analogue of pressure
 is the applied magnetic field \m{B} and its conjugate
 variable is the magnetization \m{M}. The law
 of thermodynamics becomes\footnote{For simplicity we consider only one component of the magnetic field and magnetization. When the magnetic susceptibility is isotropic this is sufficient.}
 \beq
 \alpha\equiv dU-TdS-BdM=0.
 \eeq

 The equations of state that determine the two
 dimensional surface of a magnetic material is
 known if we know \m{S(U,M)} or the Gibbs potential \m{G(B,T)}. We expect that the entropy
 will decrease with increasing  magnetization
 as the system is getting more ordered: its molecules
 are more aligned. Again by considering one mole of the paramagnetic substance, the thermodynamic law becomes
 \beq
 ds={1\over T}du-bdm,\quad b={B\over T}.
 \eeq
 where, as before, 
 \beq
 S=nRs,\quad M=nRm,\quad U=nRu
 \eeq
 and  \m{n} is the number of moles of the magnetic material.  
 
 \subsection{The Ideal Paramagnet}
 
 The magnetic analogue of an ideal gas is an ideal paramagnet, consisting of dipoles that don't interact with each other, only with an applied external field \cite{Mandl}. Thus  the internal energy of the magnets is independent of \m{m}, same as what it would have been in the absence of magnetism :
 \beq
 u=u_0(T).
 \eeq
 In these variables \m{b,m,T} , temperature `decouples' from the magnetic variables and we can treat the two degrees of freedom independent of each other.
  \beq
  ds=-bdm+{1\over T}du_0(T).
  \eeq

  Each material has   an equation of state that relates \m{b} to \m{m} . We would expect that \m{m} is an odd function of \m{b}, which tends to some finite limiting  value \m{m_0} as \m{b\to\infty}: when all the molecular magnetic moments have been aligned. A simple model that captures this idea would be
  \beq
  m=m_0f \({b m_0}\)
  \eeq
  for a constant \m{m_0} and an odd function \m{f(y)}.  The function \m{f(y)} should tend to \m{\pm 1} as \m{y\to \pm \infty}.   Also, \m{f(y)\approx y+{\rm O}(y^3)} for small \m{y} . If \m{g}  is the inverse function of \m{f}, so that  \m{g(f(y))=y},
  \beq
  b={1\over m_0}g\({m\over m_0}\).
  \eeq
  The magnetic part of the entropy would then be given by \m{{\pdr s_1\over \pdr m}=-b} so that 
\beq
s_1(m)=-\int_0^{m\over m_0}g(x)dx.
\eeq
Thus
 \beq
 s=s_1\(m\)+s_0(T)
 \eeq
 where \m{{ds_0(T)\over dT}={1\over T}{du_0(T)\over dT}}.

For example, the case of spin \m{\half} magnetic moments  
\beq
f(y)=\tanh y,\ g(x)=\half\log\({1+x\over 1-x}\),\ s_1(x)=-\half x\log\({1+x\over 1-x}\)-{1\over 2}\log[1-x^2].
\eeq
 But a simpler   choice leading to more explicit answers is 
\beq
f(y)={y\over \sqrt{1+y^2}}, \quad g(x)={x\over \sqrt{1-x^2}},\quad s_1(x)=\sqrt{1-x^2}-1.
\eeq
These two functions \m{f(x)} and \m{g(x)}  are similar enough to \m{\tanh x} and \m{{\rm arctanh}\; x} that the simpler choice already gives us a  good idea of what is going on.  
The simplest model for the non-magnetic part of the internal energy would be  that the molecules carrying the magnetic moment are oscillating around the vertices of a lattice. If there are \m{r} such oscillatory degrees of freedom, equipartition of energy gives
 \beq
 u_0(T)={r\over 2}T.
 \eeq
 Thus 
 \beq
 s_0(T)={r\over 2}\log T.
 \eeq
  
 \subsection{Curie-Weiss Theory}

Just as the van der Waals theory gives a simple model for non-ideal gases, the Curie-Weiss theory gives a simple model for ferromagnetism. Each spin is thought of as interacting with a local magnetic field that is the sum of the applied field and a constant multiple of the average of all the other spins. Thus each model of ideal paramagnet can be turned into  a  Curie-Weiss model for  ferromagnetism. The formula  for magnetization is modified to the implicit ( self-consistent) formula
\beq
m=m_0f\(m_0b+{T_c\over T}{m\over m_0}\)
\eeq
The parameter \m{T_c} ( the Curie temperature) measures the strength of the interactions of the magnets.
When \m{b=0}, the solution remains  \m{m=0} as long as \m{T<T_c}. If \m{T>T_c}, there can be spontaneous magnetization even when \m{b=0}.
By writing the equation of state as 
\beq
b={1\over m_0}g\({m\over m_0}\)-{T_c\over T} {m\over m_0^2}
\eeq
and requiring  that 
\beq
ds=-bdm+{du\over T}
\eeq
 be an exact differential, we get 
\beq
u(m,T)=-\half T_c \({m\over m_0}\)^2+u_0\(T\),\quad s=s_1\({m\over m_0}\)+s_0(T)
\eeq
where \m{u_0}  and \m{s_1} are the same functions as for  the ideal paramagnet.

We can now eliminate the parameters \m{m_0,T_c} to get  a relation between \m{s,T,b,u,m} that characterizes the Curie-Weiss models. First of all,
\beq
mb=xg(x)-x^2T_c{1\over T},\quad x={m\over m_0}.
\eeq
We eliminate \m{T_c} using ,
\beq
u=u_0(T)-\half x^2T_c
\eeq
so that 
\beq
mb=xg(x)+2{u-u_0(T)\over T}.
\eeq
If we define the functions \m{h} and \m{j} by 
\beq
h(s_1(x))=x,\quad j(y)=h(y)g(h(y))
\eeq
we get 
\beq
x=h(s-s_0(T))
\eeq
and
\beq
mb=j\(s-s_0(T)\)+2{u-u_0(T)\over T}.
\eeq
Thus the  two functions \m{s_0(T)} and \m{g(x)} characterize the mean field model, recalling 
\beq
 u_0(T)=\int_0^TTds_0(T),\quad s_1(x)=-\int_0^x g(x)dx .
\eeq

We can get an explicit form of this constraint with the simple models
\beq
u_0(T)={r\over 2}T,\quad s_1(x)=1-\sqrt{1-x^2}\Rightarrow
\eeq
\beq
g(x)= {x\over \sqrt{1-x^2} } ,\quad h(y)=\sqrt{1-(y-1)^2},\quad j(y)={1\over y}-y,
\eeq
\beq
mb={1\over s-s_0(T)}-[s-s_0(T)]+2{u-u_0(T)\over T}
\eeq
And finally
\beq
mb={1\over s-{r\over 2}\log T}-[s-{r\over 2}\log T]+2{u\over T}-r.
\eeq
\subsection{Hamilton-Jacobi Theory for Curie-Weiss Magnets}

It is simplest to use \m{T,m} as the thermodynamic variables. Then the law of thermodynamics becomes
\beq
d\Xi=-SdT+BdM,\quad \Xi=U-TS
\eeq
where \m{\Xi} is the Helmholtz Free energy. Defining \m{\Xi=nR\xi} as before
\beq
d\xi=-sdT+Tbdm, \quad s=-\({\pdr \xi\over \pdr T}\)_m,  \quad b={1\over T}\({\pdr \xi\over \pdr m}\)_T.
\eeq
The Hamilton-Jacobi equation becomes, with \m{\eta=\xi+{r\over 2T}},
\beq
\[{m\over T}{\pdr \eta\over \pdr m}-{\pdr \eta\over \pdr T}-{2r\over T^2}+2{\eta\over T}-r\]{\pdr \eta\over \pdr T}=1.
\eeq
Again we can find the general solution with two arbitrary parameters which will give the equations of state.

\section{Contact Geometry}
We can abstract out of the above a mathematical structure that captures the essence of thermodynamics.

Consider a  local patch   with co-ordinates \m{(q^0,q^1\cdots, q^n,p_1,\cdots p_n)} of a manifold with dimension \m{2n+1}. A {\em contact structure}  in this patch is  given by the one-form
\beq
\alpha\sim dq^0-p_idq^i, \cdots i=1,\cdots n
\eeq
up to multiplication by a non-zero function.  That is,  we are to regard \m{\alpha} and \m{f\alpha} (for a non-zero function \m{f})  as equivalent. It is the  vanishing of the infinitesimal variations that defines  a contact structure.

We can think of 
\beq
dq^0-p_idq^i
\eeq
 as the condition for maximizing (or minimizing )  some quantity \m{q^0} subject to the constraint that some others \m{q^i} are held constant; the \m{p_i} are the Lagrange multipliers that enforce the constraints. Such problems arise not only in statistical mechanics and thermodynamics, but also  in other areas of physics and even in economics\cite{Omohundro, microeconomics}.  

There are many other co-ordinate systems in which also the contact form  (up to multiplication by a  non-zero scalar) has the above canonical expression.  From 
\beq
f[dq^0-p_idq^i]=dQ^0-P_idQ^i
\eeq
we get the conditions 
\beq
{\pdr Q^0\over\pdr  p_i}=P_j{\pdr Q^j\over \pdr p_i},\quad -{\pdr Q^0\over \pdr q^i}+P_j{\pdr Q^j\over \pdr q^i}=p_i\left[ {\pdr Q^0\over \pdr q^0}-P_j{\pdr Q^j\over \pdr q^0}
\right]
\eeq
\beq
f={\pdr Q^0\over \pdr q^0}-P_j{\pdr Q^j\over \pdr q^0}
\eeq
Transformations \m{(q^0,q^i,p_i)\to (Q^0,Q^i,P_i)} that satisfy these conditions are called {\em Legendre transformations}.

A contact manifold is a union of co-ordinate patches such that the transformations among co-ordinates defined at the intersections are Legendre transformations. In most cases the contact manifold of interest is just  \m{R^{2n+1}} with the standard contact form above. But there are exceptions:  a superconducting circuit with a Josephson junction has \m{R^4\times S^1} as the thermodynamic phase space\cite{Ouboter}. Even on  \m{R^{3}}, there are contact structures that are not equivalent to the standard one; i.e., that need several patches to cover \m{R^3}. But it is not clear yet whether they are of interest physically. In any case, we will be content with a local description in this paper.

\subsection{ Generating Functions of  Legendre Transformations }

An infinitesimal Legendre transformation 
\beq
q^0\to q^0+tV^0,\quad q^i\to q^i+tV^i,\quad p_i\to p_i+tV_i,\quad |t|<<1
\eeq
defines a vector field 
\beq
V=V_0{\pdr \over \pdr q^0}+V^i{\pdr \over \pdr q^i}+V_i{\pdr \over \pdr p_i}
\eeq
whose components satisfy the infinitesimal version of the above condition:
\beq
{\pdr V^0\over\pdr  p_i}=p_j{\pdr V^j\over \pdr p_i},\quad -{\pdr V^0\over \pdr q^i}+p_j{\pdr V^j\over \pdr q^i}+V_i=p_i\left[ {\pdr V^0\over \pdr q^0}-p_j{\pdr V^j\over \pdr q^0}
\right]
\eeq
\beq
f={\pdr V^0\over \pdr q^0}-p_j{\pdr V^j\over \pdr q^0}
\eeq
After a little work we can see that all the components are  expressible in terms of the  single function \m{F=p_jV^j-V^0}:
\beq
V^j={\pdr F\over \pdr p_j}, V_i=-\left[{\pdr F\over \pdr q^i}+p_i{\pdr F\over q^0}\right],V^0=p_i{\pdr F\over \pdr p_i}-F.
\eeq
Thus an infinitesimal Legendre transformations is  determined by a single function \m{F}, called its {\em generating function}.

Conversely, given a function \m{F}, define the vector field 
\beq
V_F=\left[p_i{\pdr F\over \pdr p_i}-F\right]{\pdr \over \pdr q^0}-\left[{\pdr F\over \pdr q^i}+p_i{\pdr F\over \pdr q^0}\right]{\pdr \over \pdr p_i}+{\pdr F\over \pdr p_i}{\pdr \over \pdr q^i}
\eeq
The infinitesimal transformation of the contact form
\beq
\alpha=dq^0-p_idq^i
\eeq under this vector field can be calculated using
\beq
\Lie_V\alpha=d[i_V\alpha]+i_Vd\alpha.
\eeq
For us
\beq
i_{V_F}\alpha=-F,\quad d\alpha=-dp_i\wedge dq^i,\quad
 i_{V_F}d\alpha=\left[{\pdr F\over \pdr q}+p_i{\pdr F\over \pdr q^i}
\right]dq^i+{\pdr F\over \pdr p_i}dp_i.
\eeq
It follows that 
\beq
\Lie_{V_F}\alpha=-{\pdr F\over \pdr q^0}\alpha.
\eeq
Thus the vector field \m{V_F} changes the contact form only by an overall multiplication by a scalar: it leaves the contact structure unchanged. Such transformations are called Legendre transformations. 

\subsection{Characteristic Curves of a Function}
Finite Legendre transformations can be constructed by composing such infinitesimal transformations; i.e.,  by determining the integral curves of this vector field.  

Thus, a  generating function defines a one-parameter family of Legendre transformations,  determined by solving the Ordinary Differential Equations:
\beq
{dq^i\over dt}={\pdr F\over \pdr  p_i},\quad {dp_i\over dt}=-{\pdr F\over \pdr q^i}-p_i{\pdr F\over \pdr q^0},\quad {dq^0\over dt}=p_i{\pdr F\over \pdr p_i}-F.
\eeq
These are called the {\em characteristic curves}  of the Generating Function \m{F}. 

These equations are the analogues of Hamilton's equations in mechanics. But note that they are not quite the same: there are additional terms in \m{\dot  p} if   the generating function depends on \m{q^0}. Moreover,  the value of \m{F} is not always conserved:\beq
{dF\over dt}=-F {\pdr F\over \pdr q^0}.
\eeq
 But if the initial value of \m{F} is zero,  it remains   zero.

Conversely, every hypersurface can be thought of as the set of zeros of some function \m{F}. The characteristic curves of \m{F} define a dynamical system on every  hypersurface  on a contact manifold.  These curves are unchanged ( up to reparametrization) by a change 
\m{F\to \phi(F)} of the  function defining  the hypersurface. This dynamical system will play an important role in our approach to thermodynamics.

An important example of a Legendre transformation is one that interchanges a co-ordinate with its conjugate:
\beq
\tilde p_1=q^1, \tilde q^1=-p_1, \tilde p_2=p_2,\tilde q^2=q^2,\cdots
\eeq
The  characteristic curves of the function \m{F(p,q)=\half[p_1^2+(q^1)^2]} are circles: the above Legendre transformation corresponds to a rotation through \m{\pi \over 2}.
\subsection{Lagrange Brackets}

The commutator of two vector fields satisfying the conditions \m{\Lie_V\alpha=g_V\alpha, \Lie_U\alpha=g_U\alpha} is also an infinitesimal Legendre transformation. If the generating functions of \m{U} and \m{V} are \m {F} and \m{G}, what  is the  generating function of \m{[U,V]}? A straightforward calculation shows that it is given by the {\em  Lagrange bracket }
\beq
\{F,G\}=G{\pdr F\over \pdr q^0}-{\pdr G\over \pdr q^0}F+p_i\left({\pdr F\over \pdr p_i}{\pdr G\over \pdr q^0}-{\pdr F\over \pdr q^0}{\pdr G\over \pdr p_i}\right)+{\pdr F\over \pdr p_i}{\pdr G\over \pdr q^i}-{\pdr F\over \pdr q^i}{\pdr G\over \pdr p_i}
\eeq
These brackets are different from Poisson brackets of mechanics  in important ways. Although it defines a Lie algebra ( i.e., antisymmetry and  Jacobi identity are satisfied),
\beq
\{F,G\}=-\{G,F\},\quad \{F,\{G,H\}\}+\{G,\{H,F\}\}+\{H,\{F,G\}\}=0
\eeq
the Leibnitz rule of derivations is not satisfied in general:
\beq
\{F,GH\}-G\{F,H\}-\{F,G\}H=-GH{\pdr F\over \pdr q^0}\neq 0.
\eeq
Thus for example, even a constant can have non-zero brackets:
\beq
\{1,G\}=-{\pdr G\over \pdr  q^0}.
\eeq
We have  analogues of the canonical commutation relations
\beq
\{p_i,q^j\}=\delta^i_j,\quad \{q^i,q^j\}=0=\{p_i,p_j\}=\{q^0,q^i\},\quad \{q^0,p_i\}=-p_i.
\eeq
But due to violations of the Leibnitz identity these have to be used carefully. For example,
\beq
\{q^0,p_iq^i\}=0.
\eeq

\subsection{Lagrangian Submanifold}
  A Lagrangian  submanifold\cite{Arnold} of \m{M} is a submanifold of maximal dimension (i.e., \m{n}),  {\em all of whose tangent vectors are annihilated by \m{\alpha}} . As an example,  the submanifold 
  \beq
  p_1=0,\cdots p_n=0, q^0={\rm constant}
  \eeq
 with co-ordinates \m{(q^1,\cdots q^n)}  is  Lagrangian. But not every \m{n}-dimensional submanifold is  Lagrangian. For example, the \m{n}-dimensional  submanifold with co-ordinates \m{(p_1,q^1,q^3\cdots q^n)} for which 
 \beq
 q^0={\rm constant}, q^2={\rm constant}, \quad p_3=\cdots p_n=0
 \eeq
 is not Lagrangian. The independent variables on  the submanifold must have zero Lagrange bracket: they cannot include a conjugate pair. Thus a Lagrangian submanifold is the analogue of a  configuration space in mechanics.
  
 More generally,  a submanifold  is determined by a function \m{\Phi}, 
\beq
q^0=\Phi(q^1,\cdots , q^n).
\eeq
The remaining equations determining the submanifold are 
\beq
p_i=\Phi_i(q^1,\cdots q^n),\quad \Phi_i={\pdr \Phi\over \pdr q^i}.
\eeq
It is straightforward to verify that any infinitesimal variation within this surface satisfies \m{dq^0-p_idq^i=0}, since
\beq
dq^0={\pdr \Phi\over \pdr q^i}dq^i.
\eeq

The trivial case of a Lagrangian submanifold mentioned above corresponds to the choice \m{\Phi=} constant.

The same  Lagrangian submanifold can have different descriptions, as we can use different canonical co-ordinate systems on the thermodynamics phase space. These different descriptions are related by Legendre transformations.

\section{The Geometry of Thermodynamics }
Now we turn to the formulation of thermodynamics in terms of contact geometry.  As noted earlier, if a substance has \m{n}  degrees of freedom, its thermodynamic phase space is \m{2n+1} dimensional. The first law  can be see as defining  a contact structure on the thermodynamic phase space:
\beq
\alpha\equiv dq^0-p_idq^i=0.
\eeq
We can, as a first step,  think of the `co-ordinates' \m{q^i} as extensive variables and the `conjugate momenta' \m{p_i} as the intensive variables. However as noted above, a Legendre transformation can mix these up.  Having a picture that  does not depend on the choice of co-ordinates in the thermodynamic phase space can be very useful in clarifying complicated thermodynamic relations.
This is similar to the situation with canonical transformations in mechanics which allow momenta and co-ordinates to be mixed.

n the case of a gas, some of these equivalent descriptions, related by Legendre transformations,  are
 \beqs
 \alpha&=&dU-TdS+PdV\cr
  &=&d\Xi+SdT+PdV,\quad \Xi=U-TS\cr
  &=&dG+SdT-VdP,\quad G=U-TS+PV\cr
  &\sim&dS-{1\over T}dU-{P\over T}dV.
 \eeqs
\m{\Xi} is the {\em Helmholtz Free Energy} which is the convenient quantity to study a gas at constant temperature and volume. The {\em Gibbs Free Energy} \m{G} is useful to understand  a gas at constant pressure and temperature.

\subsection{Equations of State\m{\equiv}Lagrangian Submanifold}

The equations  of state of a substance define a surface of dimension \m{n} in the thermodynamic phase space.  Thus,  among the \m{2n+1} thermodynamic variables, there must be \m{n+1}  equations of state.  
But it cannot be any \m{n}-dimensional submanifold: {\em the equations of state of a substance must define a Lagrangian submanifold}. This is simply the condition that {\em  any infinitesimal change in the state of the substance must  satisfy the first law}. That is, any tangent vector field will annihilate \m{\alpha}.

Once the `fundamental relation'  giving \m{q^0=\Phi(q^1,\cdots q^n)} is given, the remaining \m{n}  equations of state follow by differentiation as above.  As an example, for a gas, the internal energy  as a function \m{u(s,v)} of the other extensive variables give the other two equations of state:
\beq
 T=\left({\pdr u\over \pdr s}\right)_v,\quad P=\left({\pdr u\over \pdr v}\right)_s
\eeq
Thus each substance has a Lagrangian submanifold corresponding to it. But the same Lagrangian can have many descriptions in different canonical co-ordinate systems. The advantage of the geometric point of view is that it allows us to choose variables according to the convenience of the physical problem. Often,  by a Legendre transformation we will be able to simplify equations that need to be solved. See below for examples.

\subsection{Family of Substances\m{\rightarrow} Hamiltonian}
A family of substances  which have similar equations of state (e.g., van der Waals gases with various values of the parameters \m{a,b} ) can be described by allowing the function \m{\Phi} to also depend on \m{n} parameters \m{a_1,\cdots a_n}:
\beq
q^0=\Phi(q^1\cdots q^n|a_1,\cdots a_n).
\eeq
These parameters could be quantities such as  critical pressure and temperature.

We require the  function defining  such a family to satisfy a non-degeneracy condition:
\beq
\det {\pdr^2\Phi\over \pdr q\pdr a}\neq 0.
\eeq
The remaining equations determining the submanifold are 
\beq
p_i=\Phi_i(q^1,\cdots q^n|a_1,\cdots a_n),\quad  \Phi_i={\pdr \Phi\over \pdr q^i}
\eeq
Given the non-degeneracy condition, we can  eliminate the parameters \m{a_1\cdots a_n} from the above \m{n+1}  equations to get a single relation among the thermodynamic variables:
\beq
F(q^0,q^1\cdots q^n,p_1,\cdots p_n)=0
\eeq
We will see that this function \m{F(q,p)} defines a dynamics on this hypersurface; by analogy to mechanics,  we call this the {\em hamiltonian} of this family of substances. 

 There are some important differences from mechanics because the physics is determined by the set of zeros of \m{F}. A  change of the hamiltonian \m{F\to \phi(F)} by an invertible function \m{\phi:R\to  R} will leave the surface unchanged and the physics will be the same as well. However, adding a constant to \m{F} {\em will} change the physics, unlike in mechanics.

Thus the thermodynamics of a family of materials is described by a contact manifold \m{M,[\alpha]} and a hyper-surface \m{F(q,p)=0} of co-dimension one on it.

Given one member of the family ( e.g., ideal gas or paramagnet) is there a mathematical  way of guessing the whole family? Unfortunately the answer is no. Such a generalization is essentially a postulate about on the macroscopic effects of the  underlying microscopic system. Some physical input is needed to make the correct generalization. The arguments of van der Waals and Curie-Weiss were great theoretical leaps at a time when very little was known  about molecules.  Given a microscopic theory, we can in principle derive \m{\Phi(q|a)} as an `effective field theory'.  In the absence of a microscopic theory, some mix of experimental information and physical intuition is needed.

Why do we define a family of substances to have exactly \m{n} parameters? With less than that many parameters, we are not talking of a `generic'  member. The non-degeneracy condition above cannot be satisfied and we would get a surface of co-dimension greater then one: it would not be defined by the vanishing of a single hamiltonian. On the other hand, with more than \m{n} parameters, we have the opposite problem: they cannot all be independent, as the surface is living in an ambient space of dimension only \m{2n+1}. Thus in a sense additional parameters are `irrelevant': they should not be important for the macroscopic dynamics of the system.  Indeed in can der Waals theoy, exactly two parameters \m{a,b}  give a good description of the non-ideal behavior of many gases.

\subsection{The Hamilton-Jacobi Equation of a Family}
Given the generating function \m{F(q,p)}  of a family of substances, we can  recover the equations of state of each member of the family by solving the first order partial differential equation 
\beq
F\(q^0,q^1\cdots q^n, {\pdr \Phi\over \pdr q^1} , \cdots {\pdr \Phi\over \pdr q^n}\)=0
\eeq
The complete integral\cite{CourantHilbert2}  of this equation will depend on  \m{n} parameters \m{a_1,\cdots a_n}.  The different equations of state of the family members are given  materials by different choices of  \m{a_1,\cdots a_n}. It is a fundamental tenet of Hamilton-Jacobi theory that  {\em solving a first order PDE is equivalent to solving a system of ODEs for   the characteristic curves.}

\subsection{The Dynamics of Thermodynamics}
 Thus, given the generating function \m{F(q,p)} of a family of substances, we get a dynamics on the hypersurface \m{F(q,p)=0}. We get\footnote{The equations in Ref.\cite{CourantHilbert2} don't have the term proportional to \m{F}. On the surface  \m{F=0} our equations agree with theirs.
}\cite{CourantHilbert2}
\beq
{dq^i\over dt}={\pdr F\over \pdr  p_i},\quad {dp_i\over dt}=-{\pdr F\over \pdr q^i}-p_i{\pdr F\over \pdr q^0},\quad {dq^0\over dt}=p_i{\pdr F\over \pdr p_i}-F.
\eeq
The parameters characterizing  the material \m{a_1\cdots a_n} are given by  the initial conditions of this dynamics. If we eliminate the `time' variable we can get the complete integral of the Hamilton-Jacobi equation which in turn gives the equations of state.

 It is tempting to speculate that this dynamics is related to the renormalization group evolution in quantum field theory:  the thermodynamic phase space 
is  `theory' space, the coupling constants being the co-ordinates on it.
Partial support for this conjecture is provided by the fact that the `time' variable of our dynamics is related to ( but is not always the same as)  entropy. 

\subsection{ The Characteristic Curves of the van der Waals gases}
 As an example,consider the case of the van der Waals family:
 \beq
 F(s,u,v,p_u,p_v)= \[ vp_v-up_u+{3\over 2}\]^2 p_u^{3}-{27\over 8}v^2e^{-2s}.    
 \eeq
  The characteristic ODEs look quite formidable in these variables.  There must be a change of variables in which the equations are simple to solve. Put first

\beq
\tilde q^0=s+{3\over 2}\log v,\quad {\tilde q}^1=v,\quad {\tilde q}^2=uv
\eeq
to  simplify the formulas. The conjugates are given by comparing 
\beq
ds-p_udu-p_vdv=dq^0-\tilde p_1d{\tilde q}^1-\tilde p_2d{\tilde q}^2\Rightarrow \tilde p_1=p_v+{3\over 2v}-{u\over v}p_u,\quad \tilde p_2={p_u\over v}
\eeq 

Then (after factoring out  an overall \m{v^5e^{-2{\tilde q^0}}})  the hypersurface can  be described by the vanishing of 
\beq
\tilde F(\tilde q^0,\tilde {\tilde q}^1,\tilde {\tilde q}^2,p_1,\tilde p_2)=e^{2\tilde q_0}\tilde p_1^2\tilde p_2^3-{27\over 8}.
\eeq
Now we make one more change of variables
\beq
p_1=e^{{2\over 5}\tilde q^0}\tilde p_1,\quad p_2=e^{{2\over 5}\tilde q^0} \tilde p_2
\eeq
and choosing \m{q^0, q^1,q^2} such that 
\beq
\alpha\sim dq^0-p_1dq^1-p_1dq_2,\Rightarrow q^0={5\over 2}e^{{2\over 5}\tilde q^0},\quad q^1={\tilde q}^1,\quad q^2={\tilde q}^2,
\eeq
\beq
 \tilde F=p_1^2p_2^3-{27\over 8}.
\eeq
Because \m{\tilde F} is independent of \m{q^0,q^1,q^2} in these variables, \m{p_1,p_2} are conserved quantities, and \m{q^0,q^1,q^2}  depend on `time' \m{t} linearly: these are the `normal co-ordinates' of the characteristic equations. Moreover from the homogeneity of \m{{\tilde F}+{27\over 8}},
\beq
{d q^0\over dt}=5{27\over 8}
\eeq
\beq
{dq^1\over dt}=2p_1p_2^3,\quad {dq^2\over dt}=3p_1^2p_2^2
\eeq

Transforming back to the original notation, we see that \m{v=q^1} and \m{uv=q^2} evolve linearly in `time' t.:
\beq
v=q^1_0+2p_1p_2^3t,\quad uv=q_0^2+3p_1^2p_2^2t, \quad q^0=5{27\over 3}t
\eeq
If we identify the constant of integration
\beq
q^1_0=b,\quad q^2_0=-a
\eeq
 and eliminate \m{t} we get 
 \beq
 {uv+a\over v-b}={3\over 2}{p_1\over p_2}={3\over 2}{\tilde p_1\over \tilde p_2}={3\over 2}{vp_v-up_u+{3\over 2}\over p_u}
 \eeq
 Transforming to the original intensive variables,we get one form of the  equation of state:
\beqs
{uv+a\over v-b}&=&{3\over 2}{vP\over T}-u{1\over T}+{3\over 2}\over {1\over T}\cr
&=&{3\over 2}[vP-u+{3\over 2}T]
\eeqs
Others forms follow similarly.

What is the  meaning of the `time' variable in this evolution, in terms of the original variable? We get (apart from a constant fixing  the origin of \m{t})
\beq
t={8\over 5\times 27}q^0 ={4\over 27}v^{3\over 5}e^{{2\over 5}s}.
\eeq
The characteristic curves  do not change if we replace \m{t} by any monotonic function of itself. Thus we can think of \m{s+{3\over 2}\log v} equivalently  as the time variable. 

The thermodynamic `time' variable is thus related to entropy, but is not exactly the same as it. Nevertheless, it seems reasonable to think of it as a measure of the  coarse-graining of the underlying microscopic system.
\section{Thermodynamics of a Black-hole}
 
It is well-known\cite{Wald}  that a black hole has an entropy proportional to  the area of its horizon. Einstein's equation imply that the total area of horizons cannot decrease as long as the source of gravity satisfies the positive energy condition (a weak form is sufficient).

 The first indication of an analogy  between  thermodynamics  and black holes  was found  in the work of  Christodoulu and Ruffini \cite{ChristodouluRuffini} who were analyzing the  Penrose process \cite{Penrose} of extracting energy from black holes. Beckenstein  \cite{Beckenstein} and Hawking\cite{Hawking}  showed  that this  was more than a mere analogy:  black holes must  have a temperature and an entropy when quantum effects are taken into account.  For a recent review including a  historical prespective, see Ref. \cite{Damour}. 

Strominger and Vafa \cite{StromingerVafa} have shown how to account for black hole entropy from string theory. Other approaches to involving  loop quantum gravity\cite{Ashtekar}, non-commutative geometry \cite{Ansoldi}  are being vigorously pursued as well.

  For simplicity let us restrict ourselves to  spherically symmetric black holes. The metric can be brought to  the form
\beq
ds^2=V(r)dt^2-{dr^2\over V(r)}-r^2d\Omega^2.
\eeq
The horizon occurs when   \m{V(r)=0}. The temperature of this  horizon is proportional to the acceleration of the null Killing vector, which works out to 
\beq
\kappa=\half V'(r_0)
\eeq
The area of the horizon  is, of course, 
\beq
A=4\pi r_0^2.
\eeq
Einstein's equation of motion for the metric imply that any infinitesimal variation of these quantities are related by an analogue of the thermodynamic law:
\beq
dM={1\over 8\pi}\kappa dA
\eeq
where \m{M} is the mass of the black hole multiplied by Newton's constant. ( In units where \m{c=1}, this has dimensions of length.)
Thus \m{A} is corresponds to entropy,  temperature to \m{\kappa\over 8\pi} and \m{M} to internal energy. This  model of a black hole has one thermodynamic degree of freedom.

\subsection{The Schwarzschild black hole}

The simplest case,  the Schwarzschild black hole, has
\beq
V(r)=1-{2M\over r}
\eeq
It has the two equations of state
\beq
\kappa={1\over 4M},\quad A=16\pi M^2
\eeq
Note that 
\beq
{\kappa\over 8\pi} ={\pdr M\over \pdr A}=\left({\pdr A\over \pdr M}\right)^{-1}
\eeq
as needed.

These equations of state are analogous to the ideal gas laws we discussed earlier. They describe a curve in the three dimensional thermodynamic phase space \m{(M,A,\kappa)}. In order to find a Hamilton-Jacobi theory, we must depart from this ideal case by introducing some parameter the deforms  Einstein's equation.

\subsection{Schwarzschild-Anti de Sitter Family}
The choice of such a deformation is a question of physics; 
the most obvious one  physically is  a cosmological constant.

 If the cosmological constant is positive, even in the absence of mass there is a horizon; a black hole will have two horizons, with associated temperatures and entropies.  Since our aim now is to keep things as simple as possible, we will  choose the cosmological constant to be negative. 
\beq
\Lambda=-{3\over l^2}.
\eeq
In the absence of mass the solution to Einstein's equations
\beq
R_{\mu\nu}-\half g_{\mu\nu}=\Lambda g_{\mu\nu}
\eeq
 is the AdS ( Anti de Sitter) metric
\beq
ds^2=\left(1+{r^2\over l^2}\right)dt^2-{dr^2\over 1+{r^2\over l^2}}-r^2d\Omega^2;
\eeq
i.e., 
\beq
V(r)=1+{r^2\over l^2}.
\eeq
This metric has no horizon as \m{V(r)} never vanishes.

 The spherically symmetric solution that has a horizon is 
\beq
ds^2=V(r)dt^2-{dr^2\over V(r)}-r^2d\Omega^2
\eeq
where
\beq
V(r)=1-{2M\over r}+{r^2\over l^2}.
\eeq
The horizon radius \m{r_0} is in 1-1 correspondence with the mass:
\beq
2M=r_0+{r_0^3\over l^2},
\eeq
since they are both positive. The area of the horizon is 
\beq
A=4\pi r_0^2.
\eeq
and the temperature is \m{\kappa\over 8\pi}, where
\beq
\kappa= \half V'(r_0)={M\over r_0^2}+{r_0\over l^2}.
\eeq
We can now eliminate the cosmological constant to get the surface in the thermodynamic phase space describing the Schwarzschild-AdS black holes:
\beq
\left[3M-{\kappa A\over 4\pi}
\right]=\sqrt{A\over 4\pi}.
\eeq
 and the Hamilton-Jacobi equation\beq
\left[3M-2 A{\pdr M\over \pdr A}
\right]=\sqrt{A\over 4\pi}.
\eeq
In this family, the Schwarzschild black hole has  the largest entropy for a given mass. In this sense it is a kind of `ground-state'.
The solution of this differential equation depends on one constant of integration, which is the cosmological constant. 

\subsection{The Characteristic Curves of  a Black hole}

The  above equation give the hamiltonian 
\beq
F(M,A,p_A)=\left(3M-2Ap_A
\right)-\sqrt{A\over 4\pi}
\eeq
Again, a Legendre transformation will simplify the dynamics. Put 
\beq
A=4\pi e^{2q}, q^0=e^{-3q}M.
\eeq
Then 
\beq
dq^0-pdq=0\Rightarrow p=e^{-3q}[2Ap_A-3M].
\eeq
After factoring out an overall factor of \m{-e^{3q}} the Hamiltonian of the S-AdS black hole becomes:
\beq
\tilde F(q^0,q,p)=p+e^{-2q}.
\eeq
The dynamical equations are, in these variables,
\beq
{dq\over dt}=1, \quad {dp\over dt}=2e^{-2q},\quad {dq^0\over dt}=- e^{-2t}.
\eeq
The solution with \m{F=0} is 
\beq
q=t,\quad p=-e^{-2t},\quad q^0=\half e^{-2t}+a
\eeq
where \m{a} is a constant of integration. Converting to the original variable \m{M=e^{3q}q^0}  and noting that \m{e^q=r_0} we get the equation of state
\beq
M=\half r_0+ar_0^3.
\eeq
Thus we identify the constant of integration with the cosmological constant.
\beq
a={1\over l^2}=-{1\over 3} \Lambda.
\eeq
In this picture the `time' variable of thermodynamic evolution is the logarithm of the entropy of the black hole.

In  general  a black hole  can carry also angular momentum and electric charge. The family of such black holes will be described by deformations of Einstein-Maxwell  equations at short distances.  The obvious deformation of Maxwell's theory  would be the Born-Infeld action. For the  Einstein  Lagrangian  itself there are several higher derivative terms that can be added. But the analysis gets quite complicated as we have to solve for the black hole metric. It should be interesting however.

Finally, recall that  the Hamilton-Jacobi formulation of mechanics  gives the shortest route to quantum mechanics.  We have considered elsewhere the possibility of a quantum thermodynamics\cite{quantumthermo}.
\section{Acknowledgement}
I would like to thank Anosh Joseph and A. P. Balachandran for many discussions, and commenting on the manuscript while in preparation. I am also grateful for  discussions  (some by email) with  Abhishek Agarwal, Levent Akant,  Sumit Das,  Gabriele Ferretti, Savitri Iyer, Samir Mathur,Mark Peterson,  Steve Omohundro   and Radu Roiban . This work was supported in part by the Department of Energy   under the  contract number  DE-FG02-91ER40685.

\end{document}